\documentclass [12pt]{article}
\usepackage[round]{natbib}
\usepackage {graphicx}
\usepackage{epsfig,amsmath,latexsym,amssymb}
\usepackage{bm}

\usepackage{mathabx}
\usepackage{booktabs}
\usepackage{multirow}
\newtheorem{algorithm_new}{Algorithm}[section]
\newtheorem{remark}{Remark}[section]

\begin{document}

\title{\vspace{-3.5cm} \textbf{Valuation of Barrier Options using\\ Sequential Monte Carlo}}
\author{\textbf{{Pavel V.~Shevchenko}}\\{\footnotesize{CSIRO Computational Informatics,
11 Julius Ave, North Ryde, NSW 2113, Australia}}\\
{\footnotesize{School of Mathematics and Statistics UNSW, Australia}}\\
{\footnotesize{Pavel.Shevchenko@csiro.au}}\\ \\
 \textbf{Pierre Del Moral}\\{\footnotesize{School of Mathematics and Statistics UNSW, NSW 2052, Australia}}\\
 {\footnotesize{p.del-moral@unsw.edu.au}}}

\date{\footnotesize{\textbf{Final version 24 July 2015, 1st version 17 May 2014}}}

\maketitle


\begin{abstract}
\noindent Sequential Monte Carlo (SMC) methods have successfully
been used in many applications in engineering, statistics and
physics. However, these are seldom used in financial option pricing
literature and practice. This paper presents SMC method for pricing
barrier options with continuous and discrete monitoring of the
barrier condition. Under the SMC method, simulated asset values
rejected due to barrier condition are re-sampled from asset samples
that do not breach the barrier condition improving the efficiency of
the option price estimator; while under the standard Monte Carlo
many simulated asset paths can be rejected by the barrier condition
making it harder to estimate option price accurately. We compare SMC
with the standard Monte Carlo method and demonstrate that the extra
effort to implement SMC when compared with the standard Monte Carlo
is very little while improvement in price estimate can be
significant. Both methods result in unbiased estimators for the
price converging to the true value as $1/\sqrt{M}$, where $M$ is the
number of simulations (asset paths). However, the variance of SMC
estimator is smaller and does not grow with the number of time steps
when compared to the standard Monte Carlo. In this paper we
demonstrate that SMC can successfully be used for pricing barrier
options. SMC can also be used for pricing other exotic options and also for cases with many underlying assets and additional stochastic factors such as stochastic volatility; we provide general formulas
and references.\\

\end{abstract}

\pagebreak

\section{Introduction}
\label{sec:introductionords}
Sequential Monte Carlo (SMC) methods (also referred to as particle methods)
have successfully been used in many applications in engineering,
statistics and physics for many years, especially in signal processing, state-space modelling and estimation of rare event probability. SMC method coincides with Quantum Monte Carlo method introduced as heuristic type scheme in physics by Enrico Fermi in 1948 while studying neutron diffusions.
From mathematical point of view SMC methods can be seen as mean field particle
interpretations of Feynman-Kac models. For a detailed analysis of these
stochastic models and applications, we refer to the couple of books \cite{dp-04,dp-13}, and references therein. The applications of these particle methods in mathematical finance has been started
recently. For instance, using the rare event interpretation of SMC, \cite{CarmonaFouque2009} and  \cite{DelMoraPatras2011} proposed SMC algorithm for computation of the probabilities of simultaneous defaults in large credit portfolios, \cite{TarginoPeters2015} developed SMC for capital allocation problems. There are many articles utilizing SMC for estimation of  stochastic volatility, jump diffusion and state-space price models, e.g. \cite{JohannesPolson2009} and \cite{PetersBrier2013} to name a few.
The applications of SMC methods in option pricing has been started recently by the second author in the series of articles \cite{carmona2012introduction,del2011robustness,del2012snell,del2012monte,jasra2011sequential}.
However, these methods are not widely known among option pricing practitioners and option pricing  literature.

 The purpose of this paper is to provide simple illustration and
explanation of SMC method and its efficiency. It can be beneficial
to use SMC for pricing many exotic options. For simplicity of
illustration, we consider barrier options with a simple geometric
Brownian motion for the underlying asset. SMC can also be used for
pricing other exotic options and different underlying stochastic
processes; we provide general formulas and references.

Barrier options  are widely used in trading.
The option is extinguished (knocked-out) or activated (knocked-in)
when an underlying asset reaches a specified level (barrier). A lot
of related more complex instruments such as bivariate barrier,
ladder, step-up or step-down barrier options have become very
popular in over-the-counter markets. In general, these options can
be considered as options with payoff depending upon the path extrema
of the underlying assets. A variety of closed form solutions for
such instruments on a single underlying asset have been obtained in
the classical Black-Scholes settings of constant volatility,
interest rate and barrier level. See for example
\cite{heynen1994partial},
\cite{kunitomo1992pricing},
\cite{rubinstein1991breaking}. If the barrier option is based on two
assets then a practical analytical solution can be obtained for some
special cases considered in \cite{heynen1994crossing}
and \cite{he1998double}.

In practice, however, numerical methods are used to price the
barrier options for a number of reasons, for example, if the
assumptions of constant volatility and drift are relaxed or payoff
is too complicated. Numerical schemes such as binomial and trinomial
lattices \citep{hull1993efficient,kat1995tree} or finite difference schemes \citep{dewynne1994partial} can be applied to the problem.
However, the implementation of these methods can be difficult. Also,
if more than two underlying assets are involved in the pricing
equation then these methods are not practical.

Monte Carlo (MC) simulation method is a good general pricing tool
for such instruments; for review of advanced Monte Carlo methods for barrier options, see  \cite{gobet2009advanced}. Many studies have been done to address finding
the extrema of the continuously monitored assets by sampling assets
at discrete dates. The standard discrete-time MC approach is
computationally expensive as a large number of sampling dates and
simulations are required. Loss of information about all parts of the
continuous-time path between sampling dates introduces a substantial
bias for the option price. The bias decreases very slowly as $1 /
\sqrt N $ for $N>>1$, where $N$ is the number of equally spaced
sampling dates (see \citealt{Broadie1997}, that also shows how to approximately calculate discretely monitored barrier option via continuous barrier case with some shift applied to the barrier).
Also, extrapolation of the Monte Carlo estimates to the continuous
limit is usually difficult due to finite sampling errors. For the
case of a single underlying asset, it was shown by \cite{AndersenBrotherton1996} and \cite{Beaglehole1997} that the bias can be
eliminated by a simple conditioning technique, the so-called
Brownian bridge simulation. The method is based on the simulation of
a one-dimensional Brownian bridge extremum between the sampled dates
according to a simple analytical formula for the distribution of the
extremum (or just multiplying simulated option payoff by the
conditional probability of the path not crossing the barrier between
the sampled dates); also see \citep[pp.
368-370]{glasserman2004monte}. The technique is very efficient in
the case of underlying asset following standard lognormal process
because only one time step is required to simulate the asset path
and its extremum if the barrier, drift and volatility are constant
over the time region. Closely related method of sampling underlying
asset conditional on not crossing a barrier is studied in \cite{glasserman2001conditioning}. The method of Brownian
bridge simulation can also be applied in the case of multiple
underlying assets as studied in \cite{Shevchenko2003}.
Importance sampling and control variates methods can be applied to
reduce the variance of the barrier option price MC estimator; for a
textbook treatment, see \cite{glasserman2004monte}. To
improve time discretization scheme convergence, in the case of more
general underlying stochastic processes,
\cite{giles2008improved, giles2008multilevel} has introduced a
multilevel Monte Carlo path simulation method for the pricing of
financial options including barrier options that improves the
computational efficiency of MC path simulation by combining results
using different numbers of time steps. \cite{gobet2010stopped} developed a procedure for multidimensional
stopped diffusion processes accounting for boundary correction
through shifting the boundary that can be used to improve the
barrier option MC estimates in the case of multi-asset and
multi-barrier options with more general underlying processes.

 However,
the coefficient of variation of the MC estimator grows when the
number of asset paths rejected by the barrier condition increases
(i.e. probability of asset path to reach maturity without breaching
the barrier decreases; for example, when barriers are getting closer
to the asset spot). This can be improved by SMC method that
re-samples asset values rejected by the barrier condition from the
asset samples that do not breach the barrier condition at each
barrier monitoring date. Both SMC and MC estimators are unbiased
and are converging to the true value as $1/\sqrt{M}$, where $M$ is
the number of simulations (asset paths) but SMC has smaller
variance.

This paper presents SMC algorithm and provides comparison
between SMC and MC estimators. We focus on the case of one
underlying asset for easy illustration, but the algorithm can easily be
adapted for the case with many underlying assets and with additional stochastic factors such as stochastic volatility. Note that we do not address the
error due to time discretization but improve the accuracy of the
option price sampling estimator for a given time discretization.



The organisation of the paper is as follows. Section \ref{Model_sec}
describes the model and notation. In Section \ref{fkf} we provide
the basic formulas for Feynman-Kac representation underlying
SMC method. Section \ref{MCestimators_sec}
presents SMC and Monte Carlo algorithms and corresponding option
price estimators. The use of importance sampling to improve SMC
estimators is discussed in Section \ref{ImportanceSampling_sec}.
Numerical examples are presented in Section
\ref{NumericalExamples_sec}. Concluding remarks are given in the
final section.

\section{Model}\label{Model_sec}
Assume that underlying asset $S_t$ follows risk neutral process
\begin{equation}\label{riskneutral_process}
dS_t=S_t \mu dt+S_t\sigma dW_t,
\end{equation}
where $\mu=r-q$ is the drift, $r$ is risk free interest rate, $q$ is
continuous dividend rate (it corresponds to the foreign interest
rate if $S_t$ is exchange rate or continuous dividends if $S_t$ is
stock), $\sigma$ is volatility and $W_t$ is the standard Brownian
motion. The interest rate can be function of time, and drift and
volatility can be functions of time and underlying asset. In this
paper, we do not consider time discretization errors; for
simplicity, hereafter, we assume that model parameters are
piece-wise constant functions of time.

\subsection{Pricing Barrier Option}
The today's fair price of continuously monitored knock-out barrier
option with the lower barrier $L_t$ and upper barrier $U_t$ can be
calculated as expectation with respect to risk neutral process
(\ref{riskneutral_process}), given information today at $t_0=0$ (i.e. conditional on $S_0=s_0$)
\begin{equation}\label{optionprice_expectation}
Q_C=B_{0,T}{E}\left(h(S_T)1_{\mathcal{A}_{t}}(S_{t})_{t\in
[0,T]}\right),\;B_{0,T}=e^{-\int_0^T r(\tau)d\tau},
\end{equation}
where $B_{0,T}$ is the discounting factor from maturity $T$ to
$t_0=0$; $1_{\mathcal{A}}(x)$ is indicator function equals 1 if
$x\in \mathcal A$ and 0 otherwise; $h(x)$ is payoff function, i.e.
$h(x)=\max(x-K,0)$ for call option and $h(x)=\max(K-x,0)$ for put
option, where $K$ is strike price; and $A_t=(L_t,U_t)$. All standard
barrier structures such as lower barrier only, upper barrier only or
several window barriers can be obtained by setting $L_t=0$ or
$U_t=\infty$ for corresponding time periods.

Assume that drift, volatility and barriers are piecewise constant
functions of time for time discretization $0=t_0<t_1<\cdots<t_N=T$.
Denote corresponding asset values as $S_0,S_1,\ldots,S_N$; the lower
and upper barriers as $L_1,\ldots,L_N$ and $U_1,\ldots,U_N$
respectively; and drift and volatility as $\mu_1,\ldots,\mu_N$ and
$\sigma_1,\ldots,\sigma_N$. That is, $L_1$ is the lower barrier for
time period $[t_0,t_1]$; $L_2$ is for $[t_1,t_2]$, etc. and similar
for the upper barrier, drift and volatility. If there is no lower or
upper barrier during  $[t_{n-1},t_n]$, then we set $L_n=0$ or
$U_n=\infty$ respectively.

Denote the transition density from $S_n$ to $S_{n+1}$ as
$f(S_{n+1}|S_{n})$ which is just a lognormal density in the case of
process (\ref{riskneutral_process}) with solution
\begin{equation}\label{process_sim}
S_n=S_{n-1}\exp\left((\mu_n-\frac{1}{2}\sigma^2_n)\delta
t_n+\sigma_n\sqrt{\delta t_n}Z_n\right),\quad n=1,\ldots, N,
\end{equation}
where $\delta t_n=t_n-t_{n-1}$ and $Z_1,\ldots, Z_N$ are independent
and identically distributed random variables from the standard
normal distribution.

In the case of barrier monitored at $t_0,t_1,\ldots,t_N$ (discretely
monitored barrier), the option price (\ref{optionprice_expectation})
simplifies to
\begin{equation}\label{Q-biased}
{Q}_D=B_{0,T}~E\left(h(S_N)~\prod_{n=1}^N~1_{(L_n,U_n)}(S_n)\right).
\end{equation}
It is a biased estimate of continuously monitored barrier option
$Q_C$ such that ${Q}_D\rightarrow Q_C$ for ${\delta t_n\rightarrow
0}$; see \cite{Broadie1997} that also shows how to approximately calculate discretely monitored barrier option via continuous barrier option price with some shift applied to the barrier in the case of one-dimension Brownian motion (for high dimensional case and more general processes, see \citealt{gobet2010stopped}).

In the case of continuously monitored barrier, the barrier option
price expectation (\ref{optionprice_expectation}) can be written as
\begin{eqnarray}
Q_C&=&B_{0,T}\int_{L_1}^{U_1}ds_1f(s_1|s_0)g(s_0,s_1)\cdots\int_{L_N}^{U_N}ds_N
f(s_N|s_{N-1})g(s_{N-1},s_{N})h(s_N),\label{optionprice_integral}
\end{eqnarray}
where $g(S_{n-1},S_n)$ is probability of no barrier hit within
$[t_{n-1},t_n]$ conditional on $S_n\in (L_n,U_n)$ and $S_{n-1}\in
(L_{n-1},U_{n-1})$. For a single barrier level $B_n$ (either lower
$B_n=L_n$ or upper $B_n=U_n$) within $[t_{n-1},t_n]$,

\begin{equation}\label{prob_nobarrierhit_singlebarrier_eq}
g(S_{n-1},S_n)=1-\exp\left(-2\frac{\ln(S_n/B_n)\ln(S_{n-1}/B_n)}{\sigma^2_n\delta
t_n}\right);
\end{equation}
and there is a closed form solution for  the case of double barrier
within $[t_{n-1},t_n]$
\begin{eqnarray}\label{prob_nobarrierhit_eq}
g(S_{n-1},S_n)&=&1-\sum_{m=1}^{\infty}[R_n\left(\alpha_n m
-\gamma_n,x_n\right)+R_n(-\alpha_n m+\beta_n,x_n)]\nonumber\\
&&+\sum_{m=1}^{\infty}[R_n(\alpha_n m,x_n)+R_n(-\alpha_n m,x_n)],
\end{eqnarray}
where
$$
x_n=\ln\frac{S_n}{S_{n-1}},\alpha_n=2\ln\frac{U_n}{L_n},\beta_n=2\ln\frac{U_n}{S_{n-1}},
\gamma_n=2\ln\frac{S_{n-1}}{L_n},
R_n(z,x)=\exp\left(-\frac{z(z-2x)}{2\sigma^2_n \delta t_n}\right).
$$
Typically few terms in the above summations are enough to obtain a
good accuracy (in the actual implementation the number of terms can
be adaptive to achieve the required accuracy; the smaller time step
$\delta t_n$ the less number of terms is needed). Formulas (\ref{prob_nobarrierhit_singlebarrier_eq}) and (\ref{prob_nobarrierhit_eq}) can easily be obtained from the well known  distribution of maximum and minimum of a Brownian motion (see e.g. \citealp{Borodin1996,KaratzasShreve1991}); also can be found in \cite{Shevchenko2011closedformdensity}.

The integral (\ref{optionprice_integral}) can be rewritten as
\begin{eqnarray}\label{optionprice_integral_1}
Q_C&=&B_{0,T}\int_{0}^{\infty}ds_1f(s_1|s_0)g(s_0,s_1)1_{(L_1,U_1)}(s_1)\cdots\nonumber\\
&&\int_{0}^{\infty}ds_N
f(s_N|s_{N-1})g(s_{N-1},s_N)h(s_N)1_{(L_U,U_N)}(s_N)\nonumber\\
&=&B_{0,T}~\times~E\left(h(S_N)~\prod_{n=1}^N~\left(1_{(L_n,U_n)}(S_n)g(S_{n-1},S_n)\right)\right).
\end{eqnarray}

Alternative expression for the barrier option that might provide
more efficient numerical estimate is presented by formula
(\ref{fk-widehat}) in the next section. It is not analysed in this paper and
subject of further study.

\subsection{Alternative Solution for Barrier Option}\label{AlternativeSolution_append}
The integral for barrier option price (\ref{optionprice_integral_1})
 can also be rewritten in terms of the Markov chain
$\widehat{S}_n$, starting at $\widehat{S}_0=S_0$, with elementary
transitions
\begin{equation}
\Pr\left(\widehat{S}_n\in
ds_n~|~\widehat{S}_{n-1}=s_{n-1}\right):=\frac{\Pr\left(S_n\in
ds_n~|~S_{n-1}=s_{n-1}\right)~1_{(L_n,U_n)}(s_n)}{ \Pr\left(S_n\in
(L_n,U_n)~|~S_{n-1}=s_{n-1}\right)}.
\end{equation}
We readily check that
\begin{equation}\label{widehtat-S}
\widehat{S}_n=\widehat{S}_{n-1}~\exp{\left(a_n+b_n
\widehat{Z}_n\right)}
\end{equation}
with
$$
a_n:=(\mu_n-\frac{1}{2}\sigma^2_n)\delta t_n\quad\mbox{\rm and}\quad
b_n:=\sigma_n\sqrt{\delta t_n}.
$$
In addition, given the state variable $\widehat{S}_{n-1}$,
$\widehat{Z}_n$ stands for a standard Gaussian random variable
restricted to the set
$\left(A_n(\widehat{S}_{n-1}),B_n(\widehat{S}_{n-1})\right)$, with
$$
A_n(\widehat{S}_{n-1}):=\left[\ln{\left(\frac{L_n}{\widehat{S}_{n-1}}\right)}-a_n\right]/b_n\quad
\mbox{\rm and}\quad
B_n(\widehat{S}_{n-1}):=\left[\ln{\left(\frac{U_n}{\widehat{S}_{n-1}}\right)}-a_n\right]/b_n.
$$

Let $\Phi(x):=\int_{-\infty}^x\frac{1}{\sqrt{2\pi}}e^{-y^2/2}dy$ be
the standard Normal (Gaussian) distribution function and its inverse
function is $\Phi^{-1}(\cdot)$. In this notation, we have that
\begin{eqnarray*}
\Pr\left(S_n\in (L_n,U_n)~|~S_{n-1}=s_{n-1}\right)&=&\Pr\left(Z_n\in \left(A_n(s_{n-1}),B_n(s_{n-1})\right)~|~S_{n-1}=s_{n-1}\right)\\
&=&\Phi(B_n(s_{n-1}))-\Phi(A_n(s_{n-1})).
\end{eqnarray*}
We can also simulate the transition $\widehat{S}_{n-1}\leadsto
\widehat{S}_{n}$ by sampling a uniform random variable ${\cal U}_n$
by taking in (\ref{widehtat-S})
$$
\widehat{Z}_n:=\Phi^{-1}\left[\Phi\left(A_n(\widehat{S}_{n-1})\right)+{\cal
U}_n~\left(\Phi\left(B_n(\widehat{S}_{n-1})\right)-\Phi\left(A_n(\widehat{S}_{n-1})\right)\right)\right].
$$
If we set
$$\varphi_{k-1}(s_{k-1}):=
\Pr\left(S_k\in
(L_k,U_k)~|~S_{k-1}=s_{k-1}\right)=\Phi(B_k(s_{k-1}))-\Phi(A_k(s_{k-1})),
$$
then we have that
$$
\begin{array}{l}
\displaystyle\left\{\prod_{k=1}^n1_{(L_k,U_k)}(s_k)\right\}~\left\{\prod_{k=1}^n\Pr\left(S_k\in ds_k~|~S_{k-1}=s_{k-1}\right)\right\}\\
\\
=\displaystyle\left\{\prod_{k=1}^n\Pr\left(S_k\in
(L_k,U_k)~|~S_{k-1}=
s_{k-1}\right)\right\}~\left\{\prod_{k=1}^n\Pr\left(\widehat{S}_k\in
ds_k~|~\widehat{S}_{k-1}=
s_{k-1}\right)\right\}\\
\\
=
\displaystyle\left\{\prod_{k=1}^{n}{\varphi}_{k-1}\left(s_{k-1}\right)\right\}~\left\{\prod_{k=1}^n\Pr\left(\widehat{S}_k\in
ds_k~|~\widehat{S}_{k-1}=s_{k-1}\right)\right\}
\end{array}
$$
from which we conclude that
\begin{equation}\label{fk-widehat}
Q_C=B_{0,T}~\times~E\left(h(\widehat{S}_N)~\prod_{n=1}^N~\widehat{G}_{n-1}(\widehat{S}_{n-1},\widehat{S}_n)\right)
\end{equation}
with the $[0,1]$-valued potential functions
\begin{equation}\label{def-G-widehat}
\widehat{G}_{n-1}(\widehat{S}_{n-1},\widehat{S}_n) :=
{\varphi}_{n-1}(\widehat{S}_{n-1})~g(\widehat{S}_{n-1},\widehat{S}_n).
\end{equation}
Explicitly, the option price integral becomes
\begin{eqnarray}\label{optionprice_integral_2}
Q_C&=&B_{0,T}\int_{0}^{1}dw_1(\Phi(\widetilde{U}_1)-\Phi(\widetilde{L}_1))g(s_0,s_1)\cdots\int_{0}^{1}dw_N
(\Phi(\widetilde{U}_N)-\Phi(\widetilde{L}_N))g(s_{N-1},s_N)h(s_N)\nonumber\\
&=&B_{0,T}\int_0^1\cdots\int_0^1dw_1\cdots dw_N h(s_N)\prod_{n=1}^N
(\Phi(\widetilde{U}_n)-\Phi(\widetilde{L}_n))g(s_{n-1},s_n),
\end{eqnarray}
where
\begin{eqnarray}
\widetilde{U}_n&=&(\ln(U_n/s_{n-1})-(\mu_n-\frac{1}{2}\sigma^2_n)\delta
t_n)/(\sigma_n\sqrt{\delta t_n}), \nonumber\\
\widetilde{L}_n&=&(\ln(L_n/s_{n-1})-(\mu_n-\frac{1}{2}\sigma^2_n)\delta
t_n)/(\sigma_n\sqrt{\delta t_n}),\nonumber\\
z_n&=&\Phi^{-1}[\Phi(\widetilde{L}_n)+w_n(\Phi(\widetilde{U}_n)-\Phi(\widetilde{L}_n))],
\nonumber\\
s_n&=&s_{n-1}\exp((\mu_n-\frac{1}{2}\sigma^2_n)\delta
t_n)+\sigma_n\sqrt{\delta t_n}z_n)\nonumber
\end{eqnarray}
are calculated from $w_1,\ldots,w_N$ recursively for
$n=1,2,\ldots,N$ for given $s_0$.

This alternative solution for the barrier option might provide
more efficient numerical estimate but it is not analysed in this paper.

\section{Feynman-Kac representations}\label{fkf}
In this section, we provide the basic option price formulas under Feynman-Kac
representation underlying SMC method; for detailed introduction of this topic, see  \cite{carmona2012introduction}.

\subsection{Description of the models} Given that the transition
valued sequence
$$
X_n=(S_n,S_{n+1})\qquad n=0,\ldots,N-1
$$
forms a Markov chain, the option price expectation in the case of
continuously monitored barrier (\ref{optionprice_integral_1}) can be
written as
\begin{equation}\label{fk-1}
Q_C=B_{0,T}~\times~E\left(H(X_N)~\prod_{n=0}^{N-1}G_n(X_n)\right)
\end{equation}
with the extended payoff functions
$$
H(X_N)=H(S_N,S_{N+1}):=h(S_N)$$
and the potential functions
$$
G_n(X_n)=
g(S_{n},S_{n+1})~\times~1_{(L_{n+1}U_{n+1})}(S_{n+1}),\quad
n=0,1,\ldots,N-1.
$$
These potential functions measure the chance to stay within the barriers during the interval $[t_p,t_{p+1}]$.
Equation (\ref{fk-1}) is the Feynman-Kac formula for discrete time models (see  \citealt{carmona2012introduction}) which is used to develop SMC option price estimator.

In this notation, the discretely monitored barrier option
expectation (\ref{Q-biased}) also takes the following form
\begin{equation}\label{Q-biased-2}
{Q}_D=B_{0,T}~\times~E\left(H(X_N)~\prod_{n=0}^{N-1}~\widetilde{G}_n(X_n)\right)
\end{equation}
with the indicator potential functions
$$
\widetilde{G}_n(X_n)=1_{(L_{n+1}U_{n+1})}(S_{n+1}),\quad
n=0,1,\ldots,N-1.
$$

We end this section with a Feynman-Kac representation of the
alternative formulae for barrier option expectation presented in
Section \ref{AlternativeSolution_append} by formula (\ref{fk-widehat}). In this case, if we consider
the transition valued Markov chain sequence
$$
\widehat{X}_n=(\widehat{S}_n,\widehat{S}_{n+1})\qquad
n=0,\ldots,N-1,
$$
based on modified underlying asset process $\widehat{S}_n$ given by (\ref{widehtat-S}), then we can rewrite the formula (\ref{fk-widehat}) as follows
\begin{equation}\label{fk-widehat_new}
Q_C=B_{0,T}~\times~E\left(H(\widehat{X}_N)~\prod_{n=0}^{N-1}~\widehat{G}_n(\widehat{X}_{n})\right)
\end{equation}
with the potential function $\widehat{G}_n$ defined in (\ref{def-G-widehat}). We observe that the above expression has exactly the
same form as (\ref{fk-1}) by replacing $(X_n,G_n)$ by  $(\widehat{X}_n,\widehat{G}_n)$.

Once the option price expectation is written in Feynman-Kac representation then it is straightforward to develop SMC estimators as described in the following sections.

\subsection{Some preliminary results}\label{SMC_preliminary_sec} In this section, we review
some key formulae related to unnormalized Feynman-Kac models. We provide a brief description of the evolution semigroup of Feynman-Kac measures. This section also presents some key multiplicative formulae describing the normalizing constants in terms of normalized Feynman-Kac measures. These mathematical objects are essential to define and to analyze particle approximation models. For instance, the particle approximation of normalizing constants are defined mimicking the multiplicative formula discussed above, by replacing the normalized probability distributions by the empirical measures of the particle algorithm.
We also emphasize that the bias and the variance analysis of these particle approximations are described in terms of the Feynman-Kac semigroups.
 A more
thorough discussion on these stochastic models is provided in the
monographs~\cite[Section 2.7.1 ]{dp-04}  and \cite[Section 3.2.2 ]{dp-13}.

Firstly, we observe that
(\ref{fk-1}) can be written in the following form
\begin{equation}
Q_C=B_{0,T}~\gamma_N(H)=B_{0,T}~\gamma_N(1)~\eta_N(H)
\end{equation}
with the Feynman-Kac unnormalized $\gamma_N$  and normalized $\eta_N$ measures given for any
function $\varphi$ by the formulae
\begin{equation}\label{FK_measures_eq}
\gamma_N(\varphi)=E\left(\varphi(X_N)~\prod_{n=0}^{N-1}G_n(X_n)\right)
\quad \mbox{\rm and} \quad
\eta_N(\varphi)=\gamma_N(\varphi)/\gamma_N(1).
\end{equation}
Notice that the sequence of non negative measures  $(\gamma_n)_{n\geq 0}$
satisfies for any bounded measurable function $\varphi$ the recursive linear equation
\begin{equation}\label{recursive_linear_eq}
\gamma_n(\varphi)=\gamma_{n-1}({\cal Q}_n(\varphi))
\end{equation}
with  the integral operator
\begin{equation}\label{integral_operator_eq}
{\cal Q}_n(\varphi)(x)=G_{n-1}(x)~K_{n}(\varphi)(x),
\end{equation}
where
\begin{equation}\label{MarkovChain_transitionOperator_eq}
K_{n}(\varphi)(x)=E\left(\varphi(X_n)~|~X_{n-1}=x\right)=\int
~K_n(x,dy)~\varphi(y)
\end{equation}
and $K_n(X_{n-1},dx):= \Pr\left(X_n\in dx~|~X_{n-1}\right)
$ is the Markov transition in the chain $X_n$.

We prove this claim using the fact that
\begin{eqnarray}
\gamma_n(\varphi)&=&E\left(E\left(\varphi(X_n)~\prod_{p=0}^{n-1}G_p(X_p)~|~(X_0,\ldots,X_{n-1})\right)\right)\nonumber\\
&=&E\left(E\left(\varphi(X_n)~~|~(X_0,\ldots,X_{n-1})\right)~\prod_{p=0}^{n-1}G_p(X_p)\right)\nonumber\\
&=&E\left(E\left(\varphi(X_n)~~|~X_{n-1}\right)~\prod_{p=0}^{n-1}G_p(X_p)\right)\nonumber\\
&=&E\left(G_{n-1}(X_{n-1})K_{n}(\varphi)(X_{n-1})
~\prod_{p=0}^{n-2}G_p(X_p)\right)\nonumber\\
&=&\gamma_{n-1}\left(G_{n-1}\times
K_{n}(\varphi)\right).
\end{eqnarray}

By construction, we also have that
\begin{eqnarray}
\gamma_N(1)&=&E\left(\prod_{n=0}^{N-1}G_n(X_n)\right)\nonumber\\
&=&E\left(G_{N-1}(X_{N-1})\times \prod_{n=0}^{N-2}G_n(X_n)\right)=\gamma_{N-1}(G_{N-1}).
\end{eqnarray}
This yields
\begin{equation}
\gamma_N(1)=\gamma_{N-1}(1)~\frac{\gamma_{N-1}(G_{N-1})}{\gamma_{N-1}(1)}=\gamma_{N-1}(1)~\eta_{N-1}(G_{N-1})
\end{equation}
from which we conclude that
\begin{equation}
\gamma_{N}(1)=\prod_{0\leq n<N}\eta_{n}(G_n)
\end{equation}
and therefore
\begin{equation}\label{FK_formula}
Q_C=B_{0,T}~\times~\left[\prod_{0\leq
n<N}\eta_{n}(G_n)\right]~\times~\eta_N(H),
\end{equation}
which is used for SMC estimators by replacing $\eta_n$ with its empirical approximation as described in the following sections.

\section{Monte Carlo estimators}\label{MCestimators_sec}
In this section we present MC and SMC estimators and corresponding
algorithms to calculate option price in the case of continuously and
discretely monitored barrier conditions.

\subsection{Standard Monte Carlo}
Using process (\ref{process_sim}), simulate independent asset path
realizations $\bm{S}^{(m)}=(S_1^{(m)},\ldots,S_N^{(m)})$,
$m=1,\ldots,M$. Then, the unbiased estimator for continuously
monitored barrier option price integral
(\ref{optionprice_integral_1}) is a standard average of option price payoff realisations over simulated paths
\begin{eqnarray}
\widehat{Q}^{MC}_C&=&B_{0,T}~~\frac{1}{M}\sum_{m=1}^M \left(
h(S_N^{(m)}) \prod_{n=1}^{N}
\left\{g(S^{(m)}_{n-1},S^{(m)}_n)1_{[L_n,U_n]}(S^{(m)}_n)\right\}\right)\nonumber\\
&=&B_{0,T}~~\frac{1}{M}\sum_{m=1}^M \left(
H(X_N^{(m)}) \prod_{n=0}^{N-1}
G_n(X_n^{(m)})\right)
\end{eqnarray}
with $X_n^{(m)}=(S^{(m)}_n,S^{(m)}_{n+1})$ and the unbiased estimator for discretely monitored barrier option
(\ref{Q-biased}) is
\begin{equation}
\widehat{Q}^{MC}_D=B_{0,T}~~\frac{1}{M}\sum_{m=1}^M \left(
H(X_N^{(m)}) \prod_{n=0}^{N-1}
\widetilde{G}_n(X_n^{(m)})\right)
\end{equation}
with $X_n^{(m)}=(S^{(m)}_n,S^{(m)}_{n+1})$.

\subsection{Sequential Monte Carlo}\label{smc-ips}\label{SMCalgo_sec}
Another unbiased estimator for option price integral
(\ref{optionprice_integral_1}) can be obtained using formula (\ref{FK_formula}) via SMC method with
the following algorithm.
\begin{itemize}
\item {\bf Initial step}
\begin{enumerate}
\item ({\em proposition step}) For the initial time step, $I_0=[t_0,t_1]$, simulate $M$ independent realizations
$$X_0^{(m)}:=(S_0^{(m)},S_1^{(m)})\qquad m=1,\ldots,M$$ using process (\ref{process_sim}); these
are  referred to as $M$ (transition type) particles. Set
$$G_0(X_0^{(m)})=1_{(L_1,U_1)}(S^{(m)}_1)~\times~g(S_0^{(m)},S^{(m)}_{1})$$
for each $1\leq m\leq M$.

\item ({\em acceptance-rejection step}) Sample $M$ random and $[0,1]$-valued uniform variables $U^{(m)}_0$. The rejected transition type particles $X_0^{(m)}$ are those for
which $G_0(X_0^{(m)})<U^{(m)}_0$. The particles $X_0^{(m)}$ for
which $U^{(m)}_0\leq G_0(X_0^{(m)})$ are accepted. Notice that a
transition type particle $X_0^{(m)}$ s.t. $S_1^{(m)}\not\in
(L_1,U_1)$ is instantly rejected (since its weight
$G_0(X_0^{(m)})=0$ is null); and a transition type particle
$X_0^{(m)}$ s.t. $S_1^{(m)}\in (L_1,U_1)$ is rejected with a
probability $1-G_0(X_0^{(m)})$.

\item ({\em recycling-selection step})
Resample each rejected transition type particle $X_0^{(m)}$ by resampling its $S_1^{(m)}$ component from the discrete
distribution with  density function
\begin{equation}\label{def-q-dist}
f(s_1)=\sum_{m=1}^M\frac{G_0(X_0^{(m)})}{\sum_{k=1}^M G_0(X_0^{(k)})}~\delta({s_1-S_1^{(m)}}),
\end{equation}

where $\delta({y-y_0})$ is a point mass function centered at $y_0$ (i.e. the Dirac $\delta$-function which is zero everywhere except from $y=y_0$ and its integral over any interval containing $y_0$ is equal to one).  In
other words, when a transition type particle, say $X_0^{(r)}$, is
rejected for some index $r$ we replace it by
one of the particle  $X_0^{(m)}$ randomly chosen w.r.t. its weight $\frac{G_0(X_0^{(m)})}{\sum_{k=1}^MG_0(X_0^{(k)})}$.\\
Efficient and simple sampling of the rejected
particle from the discrete density (\ref{def-q-dist})
can be accomplished by Algorithm \ref{sampling_from_discrete_algo} in Section \ref{resampling_algo}.

At the end of the acceptance-rejection-recycling scheme, we have $M$ (transition-type) particles that we denote
$$\widetilde{X}_0^{(m)}=({S}_0^{(m)},\widetilde{S}_1^{(m)})\qquad m=1,\ldots,M.$$

\begin{remark}\label{remark_initialstep}
By definition of (\ref{def-q-dist}) we notice that
transition type particles $X_0^{(m)}$ s.t. $S_1^{(m)}\not\in
(L_1,U_1)$ have a null weight. Therefore, they cannot be
selected in replacement of the rejected ones. Moreover, the
transition type particles $X_0^{(m)}$ s.t.
 $S_1^{(m)}\in (L_1,U_1)$ with a large
probability $G_0(X_0^{(m)})$ of non hitting the barrier within $[t_0,t_1]$ are more likely to be selected (in replacement of the rejected ones).\\
\end{remark}

\end{enumerate}

\item {\bf Step $0\leadsto 1$}
\begin{itemize}
\item[a)] ({\em proposition}) For the 2nd time step, $I_1=[t_1,t_2]$ simulate $M$ independent realizations
$$X_1^{(m)}:=(\widetilde{S}_1^{(m)},S_2^{(m)})\qquad m=1,\ldots,M$$ {\bf starting from the end
points $\widetilde{S}_1^{(m)}$ of the selected transitions
$\widetilde{X}_0^{(m)}$ at the previous step,} using the process
evolution (\ref{process_sim}); these are  referred to as $M$
(transition type) particles $X_1^{(m)}$ at time $1$. Set
$$G_1(X_1^{(m)})=1_{(L_2,U_2)}(S^{(m)}_2)~\times~g(\widetilde{S}_1^{(m)},S_2^{(m)})$$
for each $1\leq m\leq M$.

\item[b)] ({\em acceptance-rejection}) Sample $M$ random and $[0,1]$-valued uniform variables $U^{(m)}_1$. The rejected transition type particles $X_1^{(m)}$ are those for
which $G_1(X_1^{(m)})<U^{(m)}_1$ and the particles $X_1^{(m)}$ for
which $U^{(m)}_1\leq G_1(X_1^{(m)})$ are accepted. That is a
transition type particle $X_1^{(m)}$ s.t. $S_2^{(m)}\not\in
(L_2,U_2)$ is instantly rejected and $X_1^{(m)}$ s.t. $S_2^{(m)}\in (L_2,U_2)$ is rejected with a
probability $1-G_1(X_1^{(m)})$.

\item[c)] ({\em recycling-selection})
Resample each rejected transition type particles $X_1^{(m)}$  by resampling its $S_2^{(m)}$ component from the discrete
distribution with the density
\begin{equation}\label{def-q-dist-1}
f(s_2)=\sum_{m=1}^M\frac{G_1(X_1^{(m)})}{\sum_{k=1}^MG_1(X_1^{(k)})}~\delta({s_2-S_2^{(m)}})
\end{equation}
using e.g. efficient and simple Algorithm \ref{sampling_from_discrete_algo} in Section \ref{resampling_algo}.

At the end of the acceptance-rejection-recycling scheme, we have $M$ (transition-type) particles denoted as
$\widetilde{X}_1^{(m)}=(\widetilde{S}_1^{(m)},\widetilde{S}_2^{(m)})$, $1\leq m\leq M$. A remark similar to Remark \ref{remark_initialstep} is also applied here: transition type particles $X_1^{(m)}$ s.t. $S_2^{(m)}\not\in (L_2,U_2)$ have a null
weight and therefore they cannot be selected in
replacement of the rejected ones. Moreover, the transition type
particles $X_1^{(m)}$ s.t.
 $S_2^{(m)}\in (L_2,U_2)$ with a large
probability $g(\widetilde{S}_1^{(m)},S_2^{(m)})$ of non hitting the barrier within $[t_0,t_1]$ are more likely to be selected (in replacement of the rejected ones).\\

 \end{itemize}

\item Repeat steps a) to c) in {\bf Step $0\leadsto 1$} for time steps $[t_2,t_3]$,\ldots, $[t_{N-1},t_N]$.

\end{itemize}

 Calculate the final unbiased option price estimator as
\begin{eqnarray}\label{SMC_estimator_cont}
{\widehat{Q}}^{SMC}_C&=&B_{0,T} \times \left[\prod_{n=0}^{N-1}
\frac{1}{M}\sum _{m=1}^M G_n(X_n^{(m)})\right]\times
\frac{1}{M}\sum_{m=1}^M H(X_N^{(m)}).
\end{eqnarray}
That is, $\eta_N(H)$ is replaced by its empirical approximation $\frac{1}{M}\sum_{m=1}^M H(X_N^{(m)})$ and $\eta_n(G_n)$ is replaced by its empirical approximation $\frac{1}{M}\sum _{m=1}^M G_n(X_n^{(m)})$ in formula (\ref{FK_formula}).

Note that $H(X_N^{(m)})=h(\widetilde{S}_N^{(m)})$, i.e. payoff at
maturity is calculated using particles $\widetilde{S}_N^{(m)}$ after
rejection-recycling at maturity $t_N=T$.
 The proof of the unbiasedness properties of these
estimators is provided in Section~\ref{unbias-sec}.

In much the same way, an unbiased estimator of ${Q}_D$ defined in
(\ref{Q-biased-2}) is given by
\begin{equation}\label{SMC_estimator_discr}
{\widehat{Q}}^{SMC}_D=B_{0,T} \times \left[\prod_{n=0}^{N-1}
\frac{1}{M}\sum _{m=1}^M
\widetilde{G}_n({X}_n^{(m)})\right]\times
\frac{1}{M}\sum_{m=1}^M H({X}_N^{(m)}),
\end{equation}
where $\left({X}^{(m)}_n\right)_{0\leq n\leq N}$, $1\leq m\leq M$ is obtained by the above algorithm
        with potential functions $(G_n)_{0\leq n\leq N}$ replaced by the indicator potential functions $(\widetilde{G}_n)_{0\leq n\leq N}$.

In both cases, it may happen that all the particles exit the barrier after some proposition stage. In this case, we use the convention that
the above estimates are null.
One way to solve this problem is to consider the Feynman-Kac
description (\ref{fk-widehat_new}) for alternative option price
expression (\ref{fk-widehat}) presented in Section \ref{AlternativeSolution_append}. In this
context, an unbiased estimator of $Q_C$ is given by
\begin{equation}\label{SMC_estimator_cont2}
\widehat{{\widehat{Q}}}^{{SMC}}_C=B_{0,T} \times
\left[\prod_{n=0}^{N-1} \frac{1}{M}\sum _{m=1}^M
\widehat{G}_n(\widehat{X}_n^{(m)})\right]\times
\frac{1}{M}\sum_{m=1}^M H(\widehat{X}_N^{(m)}),
\end{equation}
where $\left(\widehat{X}^{(m)}_n\right)_{0\leq n\leq N}$, $1\leq m\leq M$,
is obtained by the above algorithm for $\left(X^{(m)}_n\right)_{0\leq n\leq N}$ with
potential functions $(G_n)_{0\leq n\leq N}$ replaced by the potential functions $(\widehat{G}_n)_{0\leq n\leq N}$ and process for $S_n$ is replaced by process $\widehat{S}_n$ as described in Section \ref{AlternativeSolution_append}.

\begin{remark}
As we mentioned in the introduction of Section \ref{SMC_preliminary_sec}, the particle estimate in (\ref{SMC_estimator_cont}) is defined as in (\ref{FK_formula}) by replacing the normalized Feynman-Kac measures $\eta_n$ by the particle empirical approximations. Formulae (\ref{SMC_estimator_discr}), and respectively (\ref{SMC_estimator_cont2}), follow the same line of arguments based on the Feynman-Kac model (\ref{Q-biased-2}), and respectively (\ref{fk-widehat_new}).
\end{remark}

\begin{figure}[!h]
 \centerline{\includegraphics[scale=1.0]{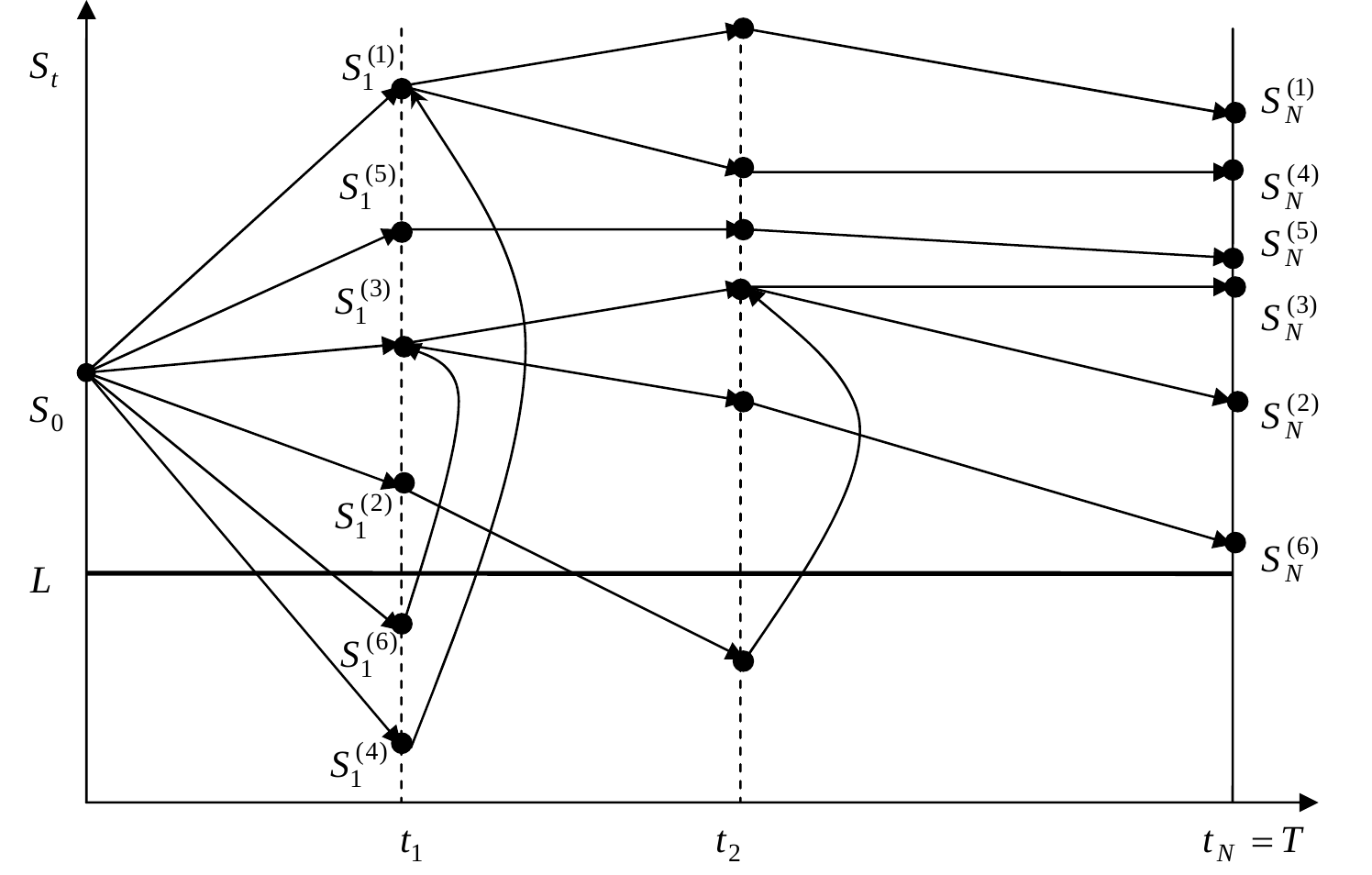}}
\caption{Illustration of Sequential Monte Carlo algorithm to
calculate barrier option with the lower barrier at level $L$.
Particle $S_1^{(4)}$ is rejected and moved to position $S_1^{(1)}$
(resampled), particle $S_1^{(6)}$ is rejected and moved to position
$S_1^{(3)}$, etc. Note that it is possible that $S_1^{(1)}$,
$S_1^{(2)}$,$S_1^{(3)}$,$S_1^{(5)}$ are also rejected in the case of
continuously monitored barrier.} \label{fig1}
\end{figure}

Figure \ref{fig1} presents an illustration of the algorithm with
$M=6$ particles. In this particular case, we simulate six particles
at time $t_1$ (starting from $S_0$). Then particle $S_1^{(4)}$ is
rejected and resampled (moved to position $S_1^{(1)}$), particle
$S_1^{(6)}$ is rejected and moved to position $S_1^{(3)}$. Then two
particles located at $S_1^{(3)}$ will generate two particles at
$t_2$, two particles located at $S_1^{(1)}$ will generate two
particles at $t_2$, etc. For each time slice including the last
$t_N$, after resampling, we have six particles above the barrier.
Note that it is possible that $S_1^{(1)}$,
$S_1^{(2)}$,$S_1^{(3)}$,$S_1^{(5)}$ are also rejected in the case of
continuously monitored barrier.

\subsection{Sampling from discrete distribution}\label{resampling_algo}
For the benefit of the reader, in this section we present efficient and simple algorithm for sampling of the rejected particles from the discrete density required during \emph{recycle-selection} step of SMC algorithm described in previous section, i.e. sampling from discrete densities (\ref{def-q-dist}) and (\ref{def-q-dist-1}).

In general, sampling of $R$ independent  random
variables $(Y^{(r)})_{1\leq r\leq R}$ from a weighted discrete probability
density function
\begin{equation}\label{ordered-stats-dist}
f(x)=\sum_{m=1}^M p_m\delta({x-x_m})
\end{equation} can be done in the usual way by the inverse distribution method. That is, $F(x)=\frac{1}{M}\sum_{m=1}^M 1_{[x_m,\infty)}(x)$ is a distribution corresponding to discrete density (\ref{ordered-stats-dist}) and $X=F^{-1}(U)$ is a sample from $F(x)$ if $U$ is from uniform (0,1) distribution. It is important to use computationally efficient method for sampling of $R$ variables. If the order of samples is not important (as in the case of recycling-selection steps of SMC algorithm in Section \ref{SMCalgo_sec}) then, for example, one can sample $(R+1)$ independent exponential random variables $({\cal E}_r)_{1\leq r\leq (R+1)}$ with unit parameter and set
\begin{equation}\label{ordered-stats}
{\cal T}_r=\sum_{1\leq s\leq r}{\cal E}_s\quad \mbox{\rm and}\quad
{\cal V}_r={\cal T}_r/{\cal T}_{R+1},\quad r=1,2,\ldots,R+1.
\end{equation}

The random variables $({\cal V}_1,\ldots,{\cal V}_R)$ calculated in such a way are the order statistics of $R$ independent random variables uniformly distributed on (0,1), which is a well known property of Poisson process, see e.g. \cite[Example 3.6.9 and Section 2.6.2]{BartoliDelMora2001} or \cite[Exercise 2.1.2]{DaleyJones2003}. Then sampling of $(Y^{(r)})_{1\leq r\leq R}$, by calculating $Y^{(r)}=F^{-1}({\cal V}_r)$, can be accomplished using the following synthetic pseudo code.

\noindent\hrulefill
\begin{algorithm_new}\label{sampling_from_discrete_algo}

~

\begin{enumerate}
\item $k=1$ and $r = 1$
\item While $r\leq R$
\begin{itemize}
\item While ${\cal V}_r<p_1+\cdots+p_k$
\begin{itemize}
\item $Y^{(r)}=x_k$
\item $r = r+1$
\end{itemize}
\item End while
\item $k= k+1$
\end{itemize}
\item End while
\end{enumerate}
\end{algorithm_new}
\noindent\hrulefill

The computational cost of this sampling scheme is linear with respect to $R$.
In particular, to simulate from the probability density (\ref{def-q-dist}) set $p_m=\frac{G_0(X_0^{(m)})}{
\sum_{k=1}^MG_0(X_0^{(k)})}$ and $x_m=S_1^{(m)}$, and
to simulate from the discrete distribution (\ref{def-q-dist-1}) set  $p_m=\frac{G_1(X_1^{(m)})}{
\sum_{k=1}^MG_1(X_1^{(k)})}$ and $x_m=S_2^{(m)}$ in (\ref{ordered-stats-dist}).

\subsection{Unbiasedness properties}\label{unbias-sec}
SMC estimator for option price (\ref{SMC_estimator_cont}) can be written as
\begin{eqnarray}
{\widehat{Q}}^{SMC}&=&B_{0,T} \times \gamma^M_{N}(1)  \times \eta^M_{N}(H)
\end{eqnarray}
with the empirical measures $\eta^M_{N}$ given by
\begin{equation}
\eta^M_{N}(H)= \frac{1}{M}\sum_{m=1}^M
H(X_N^{(m)})
\end{equation}
and  normalizing constants
\begin{equation}
\gamma^M_{N}(1)  =\prod_{p=0}^{N-1} \frac{1}{M}\sum _{m=1}^M
G_p(X_p^{(m)})=\prod_{p=0}^{N-1}\eta^M_{p}(G_p).
\end{equation}
In this notation, the $M$-particle approximations of the Feynman-Kac
measures $\gamma_N$ for any function $\varphi$ are given by
\begin{equation}
\gamma^M_{N}(\varphi):=\gamma^M_{N}(1)\times \eta^M_{N}(\varphi)
\quad\Rightarrow\quad
{\widehat{Q}}^{SMC}=B_{0,T} \times \gamma^M_{N}(H).
\end{equation}
Here, $\eta_N^M$ and $\gamma_N^M$ are particle empirical approximations of Feynman-Kac measures $\eta_N$ and $\gamma_N$ in the option price formula (\ref{FK_formula}).

The objective of this section is to show that the $M$-particle estimates ${\widehat{Q}}^{SMC}$ for continuous and discrete cases (\ref{SMC_estimator_cont}) and (\ref{SMC_estimator_discr}) are unbiased.
The unbiased property is not so obvious mainly because it is based on biased $M$-empirical measures $\eta^M_N$. It is clearly out of the scope of this study to present a quantitative analysis of these biased measure, we refer the reader to the monographs~\cite{dp-04,dp-13}, and references therein. For instance, one can prove that
\begin{equation}\label{bias_eq}
\sup_{\|\varphi\|\leq 1}\left\|E\left(\eta^M_N(\varphi)\right)-\eta_N(\varphi)\right\|\leq c(N)/M
\end{equation}
for some finite positive constant $c(N)$ whose values only depend on the time horizon $N$. That is, $\eta^M_N(\varphi)$ converges to $\eta_N(\varphi)$ as $M$ increases.
The unnormalized particle measures $\gamma_N^M$ in (\ref{SMC_estimator_cont}), (\ref{SMC_estimator_discr}), and (\ref{SMC_estimator_cont2}) are unbiased. On the other hand, the empirical measures $\eta_N^M(\varphi)$ can be expressed in terms of the ratio of two unnormalized quantities $\gamma_N^M(\varphi)$ and $\gamma_N^M(1)$. Taking into considerations the fluctuation of these unnormalized particle models, the estimate of the bias (\ref{bias_eq}) is obtained  using an elementary Taylor type expansion at the first order of this ratio.

To prove that $\gamma_N^M(H)$ is unbiased, i.e. ${\widehat{Q}}^{SMC}$ is unbiased, recall that the particles evolve sequentially using a selection and a mutation transition. Thus we have the conditional expectation formula

\begin{eqnarray}
&&E\left(  \eta^M_{N}(H)
~\left|~\left(X_0^{(m)},\ldots,X_{N-1}^{(m)}\right)_{1\leq m\leq M} \right.\right)
\nonumber\\
&&\quad =E\left(  H\left(X_{N}^{(1)}\right)
~\left|~\left(X_0^{(m)},\ldots,X_{N-1}^{(m)}\right)_{1\leq m\leq M} \right.\right)
\nonumber\\
&&\quad =\displaystyle\sum_{1\leq m\leq M}\frac{G_{N-1}(X_{N-1}^{(m)})}{
\sum_{1\leq k\leq M}G_{N-1}(X_{N-1}^{(k)})}~K_N(H)(X^{(m)}_{N-1}),
\end{eqnarray}
where $K_N$ is the Markov transition integral operator of the chain $X_n^{(m)}$, $n=1,\ldots,N-1$ defined in (\ref{MarkovChain_transitionOperator_eq}).
The weighted mixture of Markov transitions expresses the fact that the particles are selected using the potential functions before to explore the solution space using the mutation transitions.
This implies that
\begin{eqnarray}
&&E\left(  \gamma^M_{N}(H)
~\left|~\left(X_0^{(m)},\ldots,X_{N-1}^{(m)}\right)_{1\leq m\leq M} \right.\right)
\nonumber\\
&&\quad=\displaystyle\left[\prod_{p=0}^{N-1}\eta^M_{p}(G_p)\right] ~\frac{1}{N}\sum_{1\leq m\leq M}\frac{G_{N-1}(X_{N-1}^{(m)})}{\frac{1}{N}
\sum_{1\leq k\leq M}G_{N-1}(X_{N-1}^{(k)})}~K_N(H)(X^{(m)}_{N-1})
\nonumber\\
&&\quad=\left[\prod_{p=0}^{N-2}\eta^M_{p}(G_p)\right] \times \eta^M_{N-1}\left({\cal Q}_N(H)\right)
\end{eqnarray}
with the one step Feynman-Kac semigroup ${\cal Q}_N$ introduced in (\ref{integral_operator_eq}).
That is
\begin{equation}\label{expectation_FK_eq}
E\left(\gamma^M_{N}(H)
~\left|~\left(X_0^{(m)},\ldots,X_{N-1}^{(m)}\right)_{1\leq m\leq M} \right.\right)=\gamma^M_{N-1}\left({\cal Q}_N(H)\right)
\end{equation}
and therefore
\begin{equation}
E\left(\gamma^M_{N}(H)\right)=E\left(\gamma^M_{N-1}\left({\cal
Q}_N(H)\right)\right).
\end{equation}
For $N=0$, we use the convention $\prod_{\emptyset}=1$ so that
$$
\gamma^M_{0}=\eta^M_0~\Rightarrow~E\left(\gamma^M_{0}(\varphi)\right)=E\left(\eta^M_{0}(\varphi)\right)=\eta_0(\varphi)=\gamma_0(\varphi)
$$
for any function $\varphi$.
Iterating (\ref{expectation_FK_eq}) backward in time, we obtain the evolution equation of the unnormalized Feynman-Kac distributions defined in (\ref{recursive_linear_eq}). Next, for the convenience of the reader, we provide a more detailed proof of the unbiased property and we further assume that
\begin{equation}
E\left(\gamma^M_{n}(\varphi)\right)=\gamma_{n}(\varphi)
\end{equation}
at some rank $n$, for any $M\geq 1$ and any $\varphi$. In this case, arguing as above we have
\begin{equation}
E\left(\gamma^M_{n+1}(\varphi)\right)=E\left(\gamma^M_{n}\left({\cal
Q}_{n+1}(\varphi)\right)\right).
\end{equation}
Under the induction hypothesis, this implies that
\begin{equation}
E\left(\gamma^M_{n+1}(\varphi)\right)=\gamma_{n}\left({\cal
Q}_{n+1}(\varphi)\right)=\gamma_{n+1}(\varphi).
\end{equation}
This ends the proof of the unbiasedness  property of ${\widehat{Q}}^{SMC}$.
The results about standard errors of these SMC unbiased estimators can be found in e.g. \cite{cerou2011nonasymptotic}). While it goes beyond the purpose of this paper to go into details of theoretical results on the variance of empirical approximations of normalized Feynman-Kac measures, it is important to mention that the standard error of the SMC estimator is proportional to $1/\sqrt{M}$ which is the same as for the standard MC estimator. However, while for MC estimator the proportionality coefficient is easily estimated as the standard deviation of simulated asset path payoffs, for SMC estimator there is no simple expression and one has to run independent calculations of SMC estimator to estimate its standard error; numerical experiments will be presented in Section \ref{NumericalExamples_sec}.

\section{Importance sampling models}\label{ImportanceSampling_sec}
The Feynman-Kac representation formulae (\ref{fk-1}) and their
particle interpretations discussed in Section~\ref{smc-ips} are far
from being unique. For instance, using
(\ref{optionprice_integral_1}), for any  non negative probability
density functions $\overline{f}(s_n|s_{n-1})$, we also have that
\begin{eqnarray}\label{optionprice_integral_1-proba-change}
Q=&&B_{0,T}\int_{0}^{\infty}ds_1 \overline{f}(s_1|s_0)\overline{g}(s_0,s_1)1_{(L_1,U_1)}(s_1)\cdots\nonumber\\
&&\int_{0}^{\infty}ds_N
 \overline{f}(s_N|s_{N-1}) \overline{g}(s_{N-1},s_N)h(s_N)1_{(L_U,U_N)}(s_N)
 \end{eqnarray}
with the potential functions
\begin{equation}\label{potential_funcs}
\overline{g}\left(\overline{S}_{n-1},\overline{S}_n\right)={g}\left(\overline{S}_{n-1},\overline{S}_n\right)\times
\frac{f(s_n|s_{n-1})}{\overline{f}(s_n|s_{n-1})}.
\end{equation}
This yields the Feynman-Kac representation
\begin{equation}\label{fk-IS0}
Q=B_{0,T}~\times~E\left(h(\overline{S}_N)~\prod_{n=1}^N~\overline{G}_n(\overline{S}_{n-1},\overline{S}_n)\right)
\end{equation}
in terms of
the potential functions
\begin{equation}
\overline{G}_n(\overline{S}_{n-1},\overline{S}_n)=1_{(L_n,U_n)}(\overline{S}_n)
\overline{g}(\overline{S}_{n-1},\overline{S}_n)
\end{equation}
and the Markov chain $\left(\overline{S}_n\right)_{n\geq 0}$, with
\begin{equation}
\Pr\left(\overline{S}_n\in
ds_n~|~\overline{S}_{n-1}\right)=\overline{f}(s_n|\overline{S}_{n-1})~ds_n.
\end{equation}
The importance sampling formula (\ref{fk-IS0}) is rather well known. The corresponding $M$-particle consist with $M$ particles evolving, between the selection times, as independent copies of the twisted Markov chain model $\overline{S}_n$; and the selection/recycling procedure favors  transitions
$\overline{S}_{n-1}\leadsto \overline{S}_n$ that increase density ratio  ${f(\overline{S}_n|\overline{S}_{n-1})}/{\overline{f}(\overline{S}_n|\overline{S}_{n-1})}
$.

We end this section with a more sophisticated change of measure related to the payoff functions.

For any sequence of positive potential functions $(h_n)_{0\leq n\leq
N}$ with $h_N=h$, using the fact that
\begin{equation}
h(\overline{S}_N)=\frac{h_N(\overline{S}_N)}{h_{N-1}(\overline{S}_{N-1})}\times
\frac{h_{N-1}(\overline{S}_{N-1})}{h_{N-2}(\overline{S}_{N-2})}\times\ldots\times
\frac{h_{1}(\overline{S}_{1})}{h_{0}(\overline{S}_{0})}\times
h_{0}(\overline{S}_{0}),
\end{equation}
we also have that
\begin{eqnarray}\label{IS_price_eq}
Q_0&=&B_{0,T}~\times~h_{0}(\overline{s}_{0})\times~E\left(\prod_{n=1}^{N}\left(
\frac{h_{n}(\overline{S}_{n})}{h_{n-1}(\overline{S}_{n-1})}~
\overline{G}_n(\overline{S}_{n-1},\overline{S}_n)\right)\right)\nonumber\\
&=&B_{0,T}~\times~h_{0}(\overline{s}_{0})\times~E\left(\prod_{n=1}^N~\widecheck{G}_n(\overline{S}_{n-1},\overline{S}_n)\right)
\end{eqnarray}
with
\begin{equation}\label{ISpotential_eq}
\widecheck{G}_n(\overline{S}_{n-1},\overline{S}_n)=\overline{G}_n(\overline{S}_{n-1},\overline{S}_n)\times \frac{h_{n}(\overline{S}_{n})}{h_{n-1}(\overline{S}_{n-1})}.
\end{equation}
For example, for the payoff functions discussed in the option pricing model (\ref{optionprice_expectation}), we can choose
\begin{equation}\label{ex-hn}
h_N(x)=h(x)=\max(K-x,0)\quad\mbox{\rm and}\quad\forall n<N\quad h_n(x)=h(x)+1
\end{equation}
Notice that the $M$-particle model associated with the potential
functions $\widecheck{G}_n$ consists from $M$ particles evolving,
between the selection times, as independent copies of the Markov
chain $\overline{S}_n$; and the selection/recycling procedure favors
transitions $\overline{S}_{n-1}\leadsto \overline{S}_n$ that
increase the ratio
${h_{n}(\overline{S}_{n})}/{h_{n-1}(\overline{S}_{n-1})}$. For
instance, in the example suggested in (\ref{ex-hn}) the transitions
$\overline{S}_{n-1}\leadsto \overline{S}_n$ exploring regions far
from the strike $K$ are more likely to duplicate.

The choice of the potential functions (\ref{potential_funcs}) allows to choose the reference Markov chain to explore randomly the state space during the mutation transitions. The importance sampling Feynman-Kac model (\ref{IS_price_eq}) is less intrusive. More precisely, without changing the reference Markov chain, the choice of the potential functions (\ref{ISpotential_eq}) allows to favor transitions that increase sequentially the payoff function. The importance sampling models (\ref{potential_funcs}) and (\ref{ISpotential_eq}) can be combined in an obvious way so that to change the reference Markov chain and favor the transitions that increase the payoff function.

\section{Numerical results}\label{NumericalExamples_sec}
\begin{figure}[!htbp]
 \centerline{\includegraphics[scale=1.0]{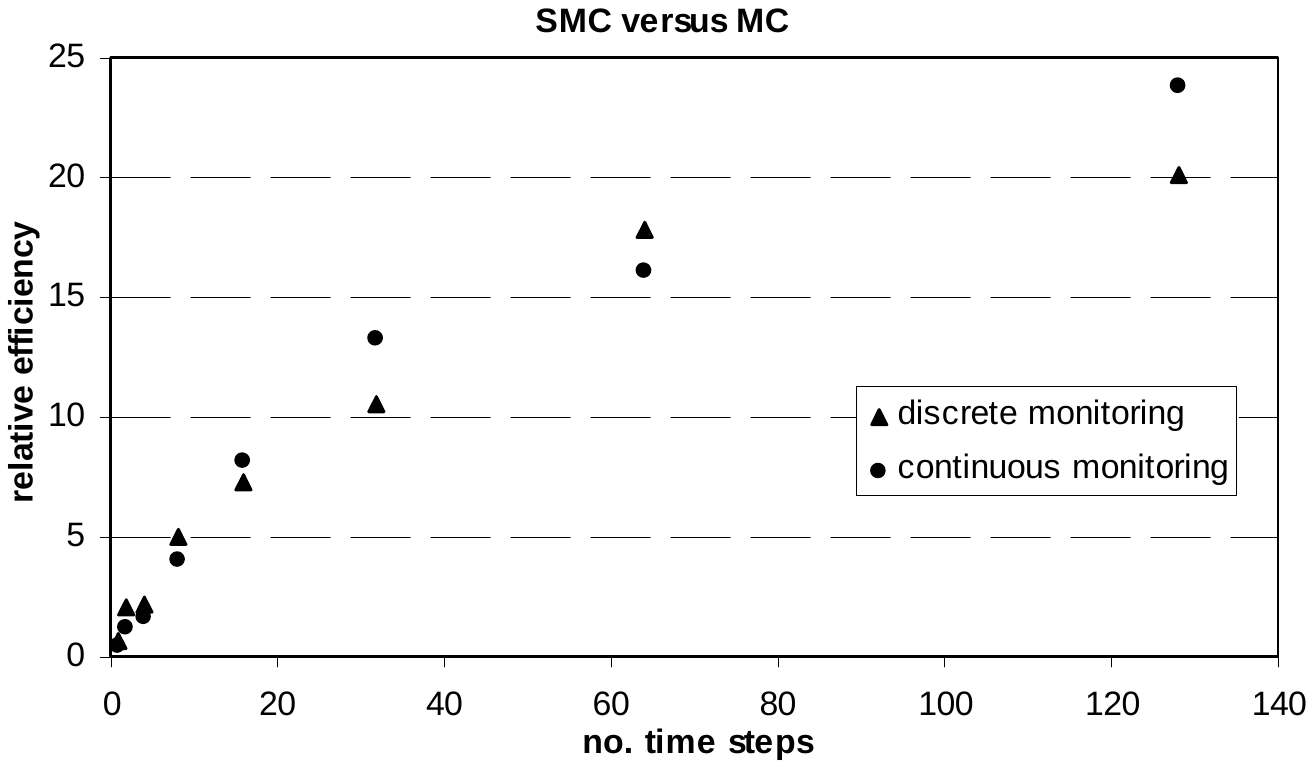}}
\caption{Relative efficiency of SMC estimator versus MC estimator
measured by coefficient $\kappa$ versus number of time steps $N$ in
the case of discretely monitored and continuously monitored barrier.
If computing time for SMC estimator is $t_{SMC}$, then the computing
time for MC estimator to achieve the same accuracy as SMC estimator
is $ \kappa \times t_{SMC}$.} \label{fig2}
\end{figure}

\begin{figure}[!htbp]
 \centerline{\includegraphics[scale=1.0]{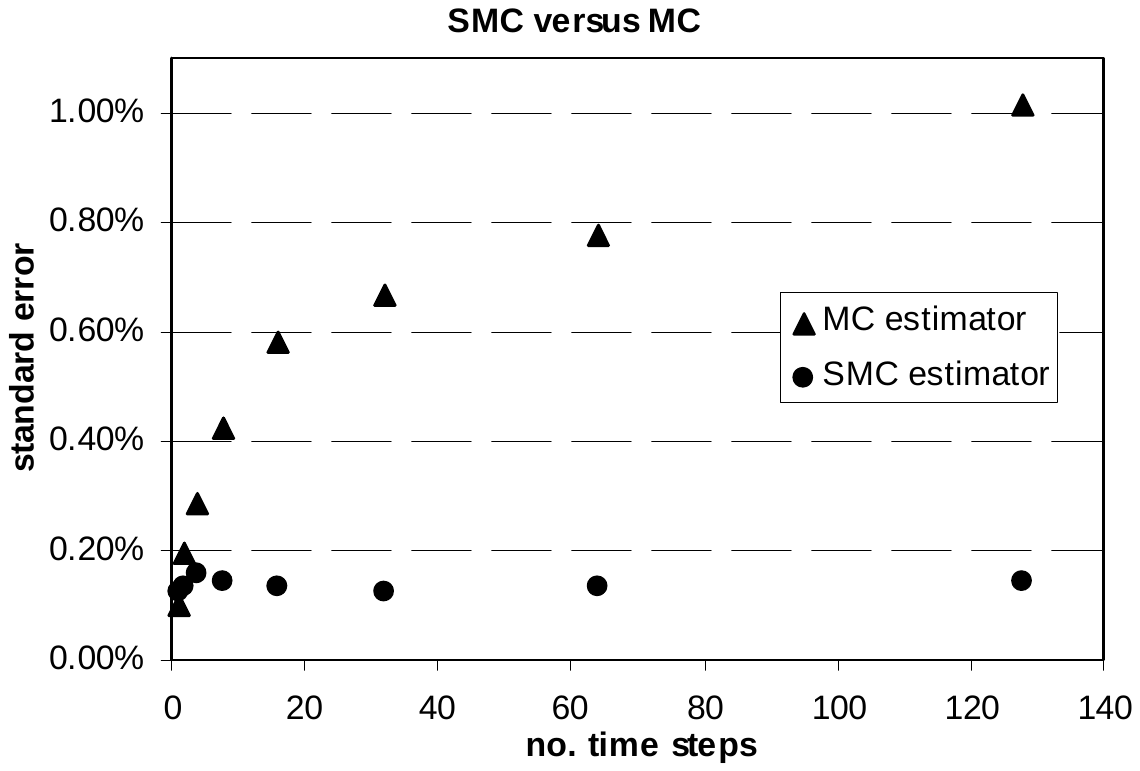}}
\caption{Relative standard error (in percent) of SMC and MC
estimators in the case of continuously monitored barrier.}
\label{fig3}
\end{figure}

Consider a simple knock-out barrier call option with constant lower
and upper barriers $L=90$ and $U=110$, strike $K=100$ and maturity
$T=0.5$ for market data: spot $S_0=100$, interest rate $r=0.1$,
volatility $\sigma=0.3$ and zero dividends $q=0$. Exact closed form
solution, SMC and standard MC estimators, standard errors of the
estimators, and estimator efficiencies for this option are presented
in Tables \ref{tab1} and \ref{tab2} and Figures \ref{fig2} and
\ref{fig3} for continuously and discretely monitored barrier cases.
We perform $M=100,000$ simulations for MC estimators and $M=100,000$
particles for SMC estimators that are repeated 50 times (using
independent random numbers) to calculate the final option price
estimates and their standard errors.

Our calculations are based on sampling at equally spaced time slices
$t_1,\ldots, t_N$. Note that we present results for
$N=(1,2,4,8,16,32,64,128)$ not to demonstrate convergence of
discretely monitored barrier to the continuous case and not to
address time discretization errors, but to illustrate and explain
the behavior of SMC that improves the accuracy of option price
sampling estimator for given time discretization. In the case of
real barrier option, the time discretization will be dictated by the
stochastic process, window barrier structure, barrier monitoring
type (e.g. continuous, daily) and market data term-structures.

For MC estimator (in the case of continuously monitored barrier) we
need to calculate conditional probability of barrier hit
(\ref{prob_nobarrierhit_eq}) between sampled dates only for asset
simulated paths that do not breach barrier condition during option
life and result in non-zero payoff at maturity, while for SMC
estimators these probabilities should be calculated for all time
steps but only for particles that appear between the barriers. Thus
direct calculation of computational effort is not straightforward.
Instead we can use the actual computing time to compare the methods
using the following facts.
\begin{itemize}
\item Computing CPU time $t_{cpu}$ is proportional to the number of
simulations $M$ in MC method (or the number of particles $M$ in
SMC).
\item Both MC and SMC estimators are unbiased. Their standard errors
are proportional to $1/\sqrt{M}$ with proportionality coefficient
for SMC different from MC (for theoretical results about variance of SMC estimators, see  \cite{cerou2011nonasymptotic}). While for MC this coefficient is easily
calculated as the standard deviation of asset path payoffs, for SMC
there is no simple expression and one has to run independent
calculations many times (i.e. 50 times in our numerical example) to
estimate standard errors of SMC estimators.
\end{itemize}
 Thus,
the squared standard error $s^2$ of an estimator is
\begin{equation}
s^2=\alpha/t_{cpu},
\end{equation}
where $\alpha$ depends on the method; i.e. $\alpha=\alpha_{MC}$ for
MC and $\alpha=\alpha_{SMC}$ for SMC that are easily found from
numerical results for $s^2$ and $t_{cpu}$ of corresponding
estimators. To compare the efficiency of the estimators we calculate
\begin{equation}
\kappa=\alpha_{MC}/\alpha_{SMC}.
\end{equation}
\emph{Interpretation  of $\kappa$ is straightforward; if computing
time for SMC estimator is $t_{SMC}$, then the computing time for MC
estimator to achieve the same accuracy as SMC estimator is $ \kappa
\times t_{SMC}$, i.e. $\kappa>1$ indicates that SMC is faster than
MC and $\kappa<1$ otherwise.}

For our specific numerical example, computing time for SMC is about
only 10\%-20\% larger than for MC in the case of discretely
monitored barrier. In the case of continuously monitored barrier,
SMC time is about twice of MC time mainly because we need to
calculate conditional probability of barrier hit
(\ref{prob_nobarrierhit_eq}) between sampled dates which is
computationally expensive in the case of double barrier. However,
standard error for SMC estimator is always smaller
 than for the MC estimator (except limiting case of $N=1$ where
barrier is monitored at maturity only when standard errors are about
the same). It is easy to see from results that SMC is superior to MC
(except the case of $N=1$). Both for discrete and continuous barrier
cases we observe that SMC efficiency coefficient $\kappa$
monotonically increases as the number of time steps $N$ increases.
The accuracy (standard error) of SMC estimator does not change much
as $N$ increases because barrier rejected asset sampled values
(particles) are re-sampled from particles between the barriers and
thus at maturity we still have $M$ particles between the barriers
regardless of $N$. Standard error of MC estimator grows with $N$
because the number of simulated paths that will reach maturity
without breaching barrier condition will reduce as $N$ increases.

It is easy to see from Table \ref{tab2} that in the case of
discretely monitored barrier, SMC efficiency $\kappa$ is about
proportional to $1/\psi$, where $\psi$ is probability of underlying
asset not hitting the barrier during option life (i.e. in this case
it is probability for the asset path to reach maturity without
breaching barrier). Note that in the case of discretely monitored
barrier, $\psi$ decreases as number of time steps $N$ increases
(i.e. less number of paths will reach maturity without breaching the
barrier as $N$ increases). In the case of continuously monitored
barrier $\psi$ does not change with $N$ (it is about 0.5\% in the
case of option calculated in our numerical example, see Table
\ref{tab1}). However, note that MC estimator for continuously
monitored barrier case is calculated by sampling asset paths through
$N$ dates and multiplying the path payoff at maturity with
conditional probabilities of not hitting the barrier between sampled
dates (\ref{prob_nobarrierhit_eq}). Thus, probability for the asset
paths to reach maturity without breaching barrier is the same as for
discrete barrier case. As a result the standard error of MC
estimator (both for discrete and continuous barrier) grows as $N$
increases.



Other numerical experiments not reported here show that efficiency
of SMC over MC improves when barriers become closer, i.e.
probability for asset path to hit the barrier increases; it is also
easy to see from results in Table \ref{tab2}. If probability of
asset path not hitting the barrier is large then performance of SMC
is about the same or slightly worse than MC. Note that our
implementation does not include any standard variance reduction
techniques such as antithetics, importance sampling and control
variates or any parallel/vector computations. The algorithm was
implemented using Fortran 90 and executed on a standard laptop
(Windows 7, Intel(R) i7-2640M CPU @ 2.8GHz, RAM 4 GB). While
computing time is somewhat subjective (i.e. depends on specifics of
our implementation), the ratio of standard errors (or ratio of
squared standard errors) of MC and SMC estimators from Tables
\ref{tab1} and \ref{tab2} strongly indicates SMC superiority over MC
having in mind that computational effort for SMC is only about
10\%-100\% larger than for MC.

\begin{table}[!htbp]
\begin{center}
\caption{Comparison MC, $\widehat{Q}_C^{MC}$, and SMC,
$\widehat{Q}_C^{SMC}$, option price estimators for continuously
monitored barriers as the number of time steps $N$ increases. Exact
price is 0.008061. Probability of the underlying asset not hitting
the barrier during option life is $\psi=0.005$.} {\footnotesize{
\begin{tabular}{ccccc}
\toprule
$N$ &  MC(stderr) &  SMC(stderr) &  $\kappa$ \\
 \midrule
1 & 0.008069(0.10\%) & 0.008074(0.12\%) & 0.39 \\
2 & 0.008059(0.19\%) & 0.008077(0.13\%) & 1.14  \\
4 & 0.008059(0.29\%) & 0.008064(0.14\%) & 1.58 \\
8 & 0.008033(0.43\%) & 0.008046(0.15\%) & 3.98 \\
16& 0.008027(0.58\%) & 0.008066(0.12\%) & 8.15 \\
32& 0.008098(0.67\%) & 0.008063(0.13\%) & 13.25 \\
64& 0.008001(0.77\%) & 0.008070(0.13\%) & 16.12 \\
128&0.007953(1.01\%) & 0.008050 (0.14\%)& 23.84 \\
 \bottomrule
\end{tabular}}}
\label{tab1}
\end{center}
\end{table}
\begin{table}[!htbp]
\begin{center}
\caption{Comparison MC, $\widehat{Q}_D^{MC}$, and SMC,
$\widehat{Q}_D^{SMC}$, option price estimators for discretely
monitored barriers as the number of time steps $N$ increases. $\psi$
is probability of the underlying asset not hitting the barrier.}
{\footnotesize{
\begin{tabular}{ccccc}
\toprule
$N$ &  MC(stderr) &  SMC(stderr) &  $\kappa$ & $\psi$\\
 \midrule
1 & 0.8225(0.11\%) & 0.8229(0.12\%) & 0.69 & 0.359\\
2 & 0.5146(0.16\%) & 0.5140(0.10\%) & 2.11 & 0.229\\
4 & 0.2985(0.16\%) & 0.2985(0.10\%) & 2.19 & 0.137\\
8 & 0.1675(0.27\%) & 0.1684(0.11\%) & 4.98 & 0.080\\
16& 0.0952(0.33\%) & 0.0957(0.11\%) & 7.25  & 0.048\\
32& 0.0568(0.44\%) & 0.0566(0.13\%) & 10.54 & 0.029\\
64& 0.0358(0.57\%) & 0.0361(0.13\%) & 17.84 & 0.019\\
128&0.0246(0.66\%) & 0.0249(0.14\%) & 20.12 & 0.013\\
 \bottomrule
\end{tabular}}}
\label{tab2}
\end{center}
\end{table}

\newpage
\section{Conclusion and Discussion}
In this paper we presented SMC method for pricing knock-out barrier
options. General observations include the following.
\begin{itemize}
\item Standard error of SMC estimator does not grow as the number of
time steps increases while standard error of MC estimator can
increase significantly. This is because in SMC, sampled asset values
(particles) rejected by barrier condition are re-sampled from asset
values between the barriers and thus the number of particles between
the barriers will not change while in MC the number of simulated
paths not breaching the barrier will reduce as the number of time
steps increases.
\item Efficiency of SMC versus standard MC improves when probability of asset path to hit the barrier
increases (e.g. upper and lower barrier are getting closer or number
of time steps increases). Typically, most significant benefit of SMC
is achieved for cases when probability of not hitting the barrier is
very small. Otherwise its efficiency is comparable to standard MC.
\item Implementation of SMC requires little extra effort when
compared to the standard MC method.
\item Both SMC and MC estimators are unbiased with standard errors proportional to $1/\sqrt{M}$,
where $M$ is the number of simulated asset paths for MC and is the
number of particles for SMC respectively; the proportionality
coefficient for SMC is different from MC.
\end{itemize}

Further research may consider development of SMC and MC for
alternative solution presented in Section
\ref{AlternativeSolution_append}. Also note that it is
straightforward to calculate knock-in option as the difference
between vanilla option (i.e. without barrier) and knock-out barrier
option, however it is not obvious how to develop
SMC estimator to calculate knock-in option directly (i.e. how to write knock-in option price expectation via Feynman-Kac representation formula (\ref{fk-1})) which is a subject of
future research. It is also worth to note that in this paper we focused on the case of one
underlying asset for easy illustration while presented SMC algorithm can easily be
adapted for the case with many underlying assets and with additional stochastic factors such as stochastic volatility.

\section*{Declaration of interest}
The authors report no conflict of interests. The authors alone are responsible for the writing of this work.

\bibliography{bibliography}
\bibliographystyle{authordate1}

\end{document}